\documentclass[11pt,a4paper]{article}
\usepackage[utf8]{inputenc}
\usepackage{cite}
\usepackage{amsmath}
\usepackage{amsfonts}
\usepackage{amssymb}
\usepackage{bbold}
\usepackage{color}
\usepackage{graphicx}
\usepackage{subcaption}
\usepackage{algpseudocode}
\usepackage{hyperref}
\usepackage{hhline}
\usepackage{comment}

\def\ep{\varepsilon}
\def\half{\frac{1}{2}}
\def\one{\mbox{1 \kern-.59em {\rm l}}}
\def\zero{\mbox{0 \kern-.59em {\rm l}}}

\def\ep{\varepsilon}
\def\half{\frac{1}{2}}

\def\K{\mathcal{K}}

\def\C{\mathcal{C}}
\def\lr#1{\left(#1\right)}
\def\slr#1{\left[#1\right]}

\def\trl#1{\textrm{Tr}\lr{#1}}
\def\strl#1{\textrm{tr}\lr{#1}}

\def\RS2{\mathbb R\times S^2_F}

\def \be  {\begin{equation}}
\def \ee  {\end{equation}}
\def \bex  {\begin{equation*}}
\def \eex  {\end{equation*}}
\def \bea {\begin{eqnarray}}
\def \eea {\end{eqnarray}}
\def \bal {\begin{align}}
\def \eal {\end{align}}

\def\no{\nonumber\\}

\begin{document}
\begin{titlepage}
\null\vspace{-62pt} \pagestyle{empty}
\begin{center}
\vspace{1truein} {\Large\bfseries
Fuzzy Onion as a Matrix Model}\\
\vskip .1in
{\Large\bfseries
~}\\
\vspace{6pt}
\vskip .1in
{\Large \bfseries  ~}\\
\vskip .1in
{\Large\bfseries ~}\\
{\large\sc Samuel Kov\'a\v{c}ik$^{[1,2]}$, Juraj Tekel$^{[1]}$}\\
\vskip .2in
[1] {\itshape Department of Theoretical Physics,\\Faculty of Mathematics, Physics and Informatics, Comenius University, Slovakia}

[2] {\itshape Department of Theoretical Physics and Astrophysics,\\Faculty of Science, Masaryk University, Brno, Czech Republic}

\vskip .1in
\begin{tabular}{r l}
E-mail:
&{\fontfamily{cmtt}\fontsize{11pt}{15pt}\selectfont samuel.kovacik@fmph.uniba.sk, juraj.tekel@fmph.uniba.sk}\\

\end{tabular}

\fontfamily{cmr}\fontsize{11pt}{15pt}\selectfont
\vspace{.8in}
\centerline{\large\bf Abstract}
\end{center}

We propose a matrix model realisation of a three-dimensional quantum space. It has an onion-like structure composed of concentric fuzzy spheres of increasing radius. The angular part of the Laplace operator is inherited from that of the fuzzy sphere. The radial part is constructed using operators that relate matrices of various sizes using the matrix harmonic expansion. As an example of this approach, we produce a numerical simulation of a scalar quantum field theory, the classical heat transfer, study the quantum mechanical hydrogen atom, and consider some analytical aspects of the scalar field theory on this space.

\end{titlepage}
\pagestyle{plain} \setcounter{page}{2}

\section{Introduction}

Some kind of quantum structure of space is expected to replace the smooth manifold structure at the order of the Planck scale \cite{doplicher,Hossenfelder:2012jw}, which is inaccessible by particle accelerators --- now or in any foreseeable future. However, the effects of such quantum structure of space have astrophysical and cosmological consequences that are already or can shortly be, within the observable range \cite{Amelino-Camelia:2008aez, Amelino-Camelia:1997ieq, Burns:2023oxn}. This is one of the possible motivations for building a model of quantum space that applies to physics in three-dimensional space.

There are various models of quantum spaces \cite{
Chamseddine:1996zu,
Ambjorn:2004qm,
Hoppe,
Madore:1991bw,
presigrosse,
seik,
patricia,
pgk15,
pg13,
Szabo:2001kg,
Steinacker:2011ix}, each of which has its benefits and drawbacks. While the construction is often motivated by instructive mathematical aspects, here, we intend to define a model closer to possible physical applications as it describes a three-dimensional space. Different models of three-dimensional quantum space are already present in the literature; two of them are close to our construction. In \cite{seik,patricia,diffstar,todorovic1,todorovic2}, a model of quantum field theory was built and studied using different modifications of star product Moyal space\footnote{\textcolor{black}{For a recent review, including discussion of gauge theories and further references, see \cite{patricia2}.}}, and in \cite{pgk15}, a quantum-mechanical model was described using an auxiliary bosonic Fock space. Both had a three-dimensional space foliated by a set of concentric fuzzy spheres of increasing radii. This structure also appeared as a black-hole solution to modified Einstein equations \cite{Schupp:2009pt}, \textcolor{black}{was identified as a solution of IKKT-like model models \cite{vuvuzela}} and was independently analyzed in the quantum-mechanical setting \cite{sch1,sch2,sch3,sch4}.

Construction presented here follows our introductory proposal \cite{corfu22}. It utilizes a similar viewpoint, but we will use explicit matrix formulation for the fuzzy spheres \cite{Hoppe,Madore:1991bw,presigrosse}, together with a different definition of the Laplace operator. The Laplacian was constructed top-to-bottom in the above-mentioned works, resulting from the natural structures used to define the space.
We use a bottom-up approach, where we define a natural way to compare field configurations on different layers and then build up derivatives and the Laplace operator.
We aim to have a model that is easily evaluated numerically, for example, using Monte Carlo (MC) methods \cite{Duane:1987de,Ydri:2015zba, Kovacik:2022kfh}. Similar methods proved to be helpful in the case of the fuzzy spheres and various types of field theories on them \cite{Ydri:2014rea, Rea:2015wta, Bosetti:2015lsa, GarciaFlores:2009hf, Panero:2006bx, Kovacik:2018thy}; the model proposed here can be treated in the same way. Our approach is based on a dual model description, either in terms of Hermitian matrices or their expansion coefficients using matrix harmonics. We also investigate this dual description, which is unsuitable for numerical methods but can provide deeper analytical insight.

We investigate several physical models defined on the fuzzy onion to test the construction, show its feasibility, and compare it with previous works. The fuzzy sphere is currently being most studied either in the context of the M-theory or to understand unresolved issues with formulating a field theory on noncommutative spaces. Therefore, the first example we elaborate on is the scalar field theory on the fuzzy onion model with $\Phi^4$-interaction term. The fuzzy spaces are usually used to investigate quantum theories but can also be used in classical settings. Their main feature --- a limited spatial resolution --- is expected to originate from Planck scale physics. Still, they might be useful as an effective description of granular materials with mesoscopic features \cite{granular}, and the limited resolution can also be used in cases where the limitation is not of a material cause but due to spatially-separated measurements, such as in meteorological models. In the present work, we investigate the classical heat transfer problem on the fuzzy onion. During the construction, the inspiration we had in mind was the $\mathbb{R}^3_\lambda$ model of three-dimensional quantum space \cite{pgk15}. The Coulomb problem was solved within it, and the energy spectrum was computed exactly as a function of the scale of space noncommutativity. Therefore, we naturally solve the problem in the fuzzy onion model and compare the results.

This paper is organized as follows: in the next section, we summarize two previous models of quantum spaces: the matrix formulation of the fuzzy sphere and bosonic operator construction of the three-dimensional space $\mathbb R_\lambda^3$. In section \ref{sec3}, we propose the matrix formulation of the fuzzy onion model. In section \ref{sec4}, we show examples of this model: an interacting scalar quantum field theory, the classical heat transfer and the quantum mechanical Coulomb problem. In section \ref{sec5}, we further analyze the Fourier picture of the model. We conclude this report with a discussion section and with technical appendices. 

\section{The fuzzy sphere $S^2_\lambda$ and the three-dimensional space $\mathbb R^3_\lambda$}

What is a sphere? The conventional definition states that it is a set of points with the same distance from a certain point or an orbit of $SO(3)$ rotation group. Another definition is that it is a space on which the infinite-dimensional representation of $su(2)$ lives. In other words, the sphere is described by an algebra of functions --- spherical harmonics --- that exist on it. 
The spherical harmonics satisfy 
\begin{equation} \label{harmonics}
     [L^{(N)}_i,[L^{(N)}_i,Y^{(N)}_{lm}]] = l(l+1)Y^{(N)}_{lm},\ \  [L^{(N)}_3, Y^{(N)}_{lm}] = m Y^{(N)}_{lm}, 
\end{equation}
where $L^{(N)}_i$ are the rotation generators that obey $[L^{(N)}_i,L^{(N)}_j] = i \varepsilon_{ijk} L^{(N)}_k$ relation and $(N)$ denotes the representation. The fuzzy sphere model relies on the existence of finite-size representations satisfying \eqref{harmonics}. These are realised as ${N \times N}$ matrices that also serve as a basis for Hermitian matrices 
\begin{equation} \label{PhiExpansion}
\Phi^{(N)} = \sum \limits_{l=0}^{N-1} \sum \limits_{m=-l}^{l} c^{(N)}_{lm} Y^{(N)}_{lm}.
\end{equation}
The superscript denotes the matrix size, and $\Phi^{(\infty)}$ corresponds to the case of the ordinary sphere, that is, $Y^{(\infty)}_{lm}$ are the spherical harmonics. Matrices $\Phi^{(N)}$ describe fields on the so-called fuzzy sphere $S^2_\lambda$.

This expansion allows us to interpret the matrices as fields on the sphere. Both of them can be expanded in terms of harmonics, and even though their numbers differ, matrices can be mapped onto fields as:
\begin{eqnarray} \label{map}
    c^{(\infty)}_{lm} &=& c^{(N)}_{lm} \text{ for } l \le N-1 \\ \nonumber c^{(\infty)}_{lm} &=& 0 \text{ otherwise}.
\end{eqnarray}

The matrices have a finite number of degrees of freedom, meaning an exact $\delta$-function cannot be constructed, and the spatial resolution is restricted. Or the same effect explained differently, invoking an upper limit on momenta, $l \le N-1$, invokes a lower limit on the shortest distinguishable lengths, $\lambda^{(N)} \propto N^{-1}$.

An integration in the case of the fuzzy sphere of unit radius is realised by taking a trace
\begin{eqnarray}\label{integral}
    \int \Phi^{(\infty)}d \Omega \rightarrow \frac{4 \pi}{N}\ \mbox{tr}_N \Phi^{(N)},
\end{eqnarray}
where we have denoted the trace over ${N\times N}$ matrices accordingly, reserving the standard notation for something different. The angular Laplace operator is defined using $L_i$ generators. For example, a $\Phi^4$-scalar field theory can be defined on the fuzzy sphere by the action
\begin{equation*}
S_N [\Phi^{(N)}] = \frac{4\pi}{N}\mbox{tr}_N \left( a\ \Phi^{(N)} {\mathcal{K}}^{(N)} \Phi^{(N)} + b\ (\Phi^{(N)})^2 + c\ (\Phi^{(N)})^4 \right),
\end{equation*}
where we have the kinetic term
\begin{align}
\mathcal{K}^{(N)} \Phi^{(N)}= [L_i^{(N)},[L_i^{(N)},\Phi^{(N)}]]\ . \label{kin_N}
\end{align}
The fuzzy sphere can also be expressed in terms of noncommuting coordinates $x_i$
\begin{equation}
[x_i,x_j]= i \varepsilon_{ijk} \frac{2r}{\sqrt{N^2-1}} x_k,
\end{equation}
where now the radius ${x^2 = r^2}$ is explicit and ${x_i = \frac{2 r}{\sqrt{N^2-1}} L_i}$. The scale of noncommutativity is set by $\lambda = \frac{r}{\sqrt{N^2-1}}$ where $r$ is the radius of the sphere. As the quantumness of space is generally predicted by theories of quantum gravity, $\lambda$ is often assumed to be of the order of Planck length. However, noncommutative spaces can appear in other contexts -- for example, the quantum Hall effect -- with a different length scale.

There is another way of constructing a space whose coordinates satisfy the relation
\begin{equation}
 [x_i,x_j]= 2 \lambda i \varepsilon_{ijk}  x_k.   
\end{equation}
A particular construction of three-dimensional quantized space $\mathbb R_\lambda^3$ has been described in \cite{pgk15} and uses two sets of auxiliary operators satisfying
\begin{equation} [\text{a}_\alpha,\text{a}^\dagger_\beta]\,=\,\delta_{\alpha\beta },\ \
[\text{a}_\alpha,\text{a}_\beta]\,=\,[\text{a}^\dagger_\alpha, \text{a}^\dagger_\beta]\,=\,0\, ,
\end{equation}
and acting on the Fock space $\cal{F}$ as 
 \begin{equation} 
\frac{(\text{a}^\dagger_1)^{n_1}\,(\text{a}^\dagger_2)^{n_2}}{\sqrt{n_1!\,n_2!}}\ |0\rangle =\ |n_1,n_2\rangle\ . 
\end{equation}
In this space, denoted $\mathbb{R}^3_\lambda$, the Cartesian and radial coordinates were defined using Pauli matrices, $\sigma^i$, as
\begin{equation}\label{op:r}
    \ x_i = \lambda \text{a}^\dagger_\alpha \sigma^i_{\alpha \beta}\text{a}_\beta\ ,\ r = \lambda \left( \text{a}^\dagger_\alpha \text{a}_\alpha + 1 \right).
\end{equation}
These satisfy $x^2 = r^2 - \lambda^2 $ which recovers $x^2 = r^2$ in the commutative limit, $\lambda \to 0$. Note that the value of $r$ is quantized as $\text{a}^\dagger_\alpha \text{a}_\alpha$ acts as the number operator on ${\cal F}$:
\begin{equation}
    r |n_1,n_2\rangle = \lambda \left( n_1+n_2+1 \right) |n_1,n_2\rangle.
\end{equation}
Let us stress that in this construction, $\lambda$ is a constant that does not change with $N$, as opposed to the construction of a single sphere with a finite radius in the large $N$ limit. We hope that which of the two notions we have in mind will be clear from the context. The space in this model can be understood as a set of concentric fuzzy spheres of increasing radius with the increment of $\lambda$. The kinetic term was defined as 
\begin{equation}
H_0 \Psi = \frac{1}{2\lambda r} [\text{a}^\dagger_\alpha ,
[\text{a}_\alpha , \Psi ]]\label{H0}.
\end{equation}
This model was used, for example, in \cite{pg13,pgk15} to study the Coulomb problem, where the spectrum was found exactly to be 
\begin{equation}
    E^{I}_{\lambda \,n} = \frac{\hbar}{m_e \lambda^2}\left(1- \sqrt{1+\lr{\frac{m_e q \lambda}{\hbar^2 n}}^2}\right), E^{II}_{\lambda \, n} = \frac{2\hbar}{m_e\lambda^2} - E^{I}_{\lambda \, n}.\label{pgk:energiesI}
\end{equation}
Note that one set of energies reproduces the hydrogen atom spectrum in the $\lambda \rightarrow 0$ limit. Then another set of solutions reflected w.r.t. the Planck scale showing a similar duality as considered recently in \cite{Nicolini:2022rlz}. In one of the later sections, we will reproduce this result numerically using the fuzzy onion model.

\section{The fuzzy onion ${\cal{O}}_\lambda$ model}\label{sec3}

In this section, we define the three-dimensional space in terms of matrices, together with derivative operators and the action of scalar field theory. The first step, glueing fuzzy spheres together, is simple, but we do it in detail to set the conventions and notations. The second step, mapping between consecutive spheres, is more demanding --- conceptually and technically. 

\subsection*{The easy part: gluing spheres together}
We will now consider $M$ concentric fuzzy spheres of increasing radius with a step $\lambda$ that form an onion-like structure. A field on each layer is described by a Hermitian matrix $\Phi^{(N)}$, the further the layer, the larger the matrix --- the innermost being described by a single element matrix $\Phi^{(1)}$. The configuration of fields on each of those layers can be described by a block-diagonal matrix 
\begin{equation} \label{psi}
    \Psi = \begin{pmatrix}
\Phi^{(1)} &  &  &\\
 & \Phi^{(2)} &  &\\
 &  & \ddots &\\
 &  &  & \Phi^{(M)}
\end{pmatrix} 
\end{equation}
of size $\frac{ M (M+1)}{2}$. The dimension of this space is
\begin{align}\label{dimd}
    d= \sum_{N=1}^{M}N^2=\frac{M (M+1)(2 M+1)}{6} \ .
\end{align}

This matrix now describes field configurations on all spherical layers. For finite $M$, this covers a fuzzy ball of radius ${R = \lambda M}$, the field $\Psi$ outside this support is taken to be vanishing. By taking ${M \rightarrow \infty}$, the layers cover the entire space $\mathbb R^3_\lambda$. On the other hand, keeping $\lambda M$ fixed while taking ${\lambda \rightarrow 0}$ leads to an ordinary continuous ball. 

The integration of fields is to be understood as summing the integration over individual layers, which can be related to a trace over this large matrix $\Psi$. More precisely, to define the integral, we recall the standard three-dimensional integration of a function $\psi$
\begin{align}
    \int \ d^3x \ \psi =\int\ r^2\ dr\int d\Omega  \ \psi 
\end{align}
and change this to a version discrete in the radial direction
\begin{align}\label{NCint}
    \sum_{N=1}^{M} (\lambda N)^2\, \lambda\, \frac{4\pi}{N}\ \mbox{tr}_N \Phi^{(N)}=
    \mbox{Tr}\lr{4\pi \lambda^2 r\ \Psi},
\end{align}
where we have defined the radial distance matrix $r$ as
\begin{equation}\label{r}
    r = \begin{pmatrix}
\lambda\ \one_{1\times1} &  &  &\\
 & 2 \lambda \ \one_{2\times2}  &  &  &\\
 &  & 3 \lambda \ \one_{3\times3} &  &\\
 &  &  & \ddots  &\\
 &  &  &  & M \lambda \ \one_{M\times M}
\end{pmatrix}
\end{equation}
and denoted the trace of the block diagonal matrices $\Psi$ by $\mbox{Tr}$ as advertised before\footnote{As a check, we can see that the integral of identity matrix yields, in the large $M$ limit, the volume of a sphere with radius $\lambda M$.}. The same formula for the integration measure was obtained in \cite{pgk15} using the definition of the fuzzy sphere as quantization of the Hopf fibration.

\textcolor{black}{Functions of the fields can be defined using the expansion series $P(\Psi) = \sum_i q_i \Psi^i$. For example,} \textcolor{black}{we can define potential for quartic scalar field theory this way and use \eqref{integral} to define the potential part of the action as follows}
\begin{equation}
V(\Psi) = 4\pi \lambda^2 \ \mbox{Tr} \left( b\ r \Psi^2 + c\ r\Psi^4 \right). 
\end{equation}

The angular part of the kinetic term can be defined in a layer-wise fashion using \eqref{kin_N} as
\begin{equation} \label{KL}
    \mathcal{K}_L \Psi = r^{-2}\begin{pmatrix}
\mathcal{K}^{(1)}\Phi^{(1)} &  &  &  &\\
 & \mathcal{K}^{(2)}\Phi^{(2)} &  &  &\\
 &  & \mathcal{K}^{(3)}\Phi^{(3)} &  &\\
 &  &  & \ddots  & \\
 &  &  &  & \mathcal{K}^{(M)}\Phi^{(M)}
\end{pmatrix}.
\end{equation}

To summarize, a field on a single fuzzy sphere is described using a matrix, so we describe a field living on multiple fuzzy spheres using a larger matrix that encompasses them all. The kinetic structure on individual layers (in the angular directions) is inherited from the single fuzzy sphere construction. Now, the nontrivial task is to connect consecutive layers.  

\subsection*{The difficult part: defining the radial derivative}

To take the derivative of $\Psi$ in the radial direction, we need to be able to compare fields on consecutive layers. However, these are expressed using matrices of different sizes and, therefore, have different degrees of freedom.

This is a crucial point we need to overcome. To do so, we can use the same trick utilised to define the map \eqref{map}. The idea is to expand the matrices -- i.e. the field configurations on the given layers -- in terms of matrix harmonics, compare coefficients that can be compared and set the rest to zero. To be exact, we define two maps, one  going one layer up, ${\cal {U}}:(N)\rightarrow (N+1)$ and one going one layer down, ${\cal {D}}:(N+1)\rightarrow (N)$, as follows
\begin{align}
    &\textrm{for }\Phi^{(N)} = \sum \limits_{l=0}^{N-1} \sum \limits_{m=-l}^l c^{(N)}_{lm} Y_{lm}^{(N)}\ , \ \Phi^{(N+1)} = \sum \limits_{l=0}^{N} \sum \limits_{m=-l}^l c^{(N+1)}_{lm} Y_{lm}^{(N+1)} \nonumber \\
    \mathcal{D}:\Phi^{(N+1)}\to\,&\Phi^{(N)}=\sum \limits_{l=0}^{N-1} \sum \limits_{m=-l}^l c^{(N)}_{lm} Y_{lm}^{(N)}\ ,\ c^{(N)}_{lm}=c^{(N+1)}_{lm} \text{ for }   l\le N-1\\
    \mathcal{U}:\Phi^{(N)}\to\,&\Phi^{(N+1)}=\sum \limits_{l=0}^{N} \sum \limits_{m=-l}^l c^{(N+1)}_{lm} Y_{lm}^{(N+1)}\ ,\
    \left\{
    \begin{array}{l}
    c^{(N+1)}_{lm}=c^{(N)}_{lm} \text{ for }   l\le N-1\\  c^{(N+1)}_{Nm}=0\end{array}\right.\ .
\end{align}
Or expressed in words: when making a matrix larger, add necessary coefficients, all with zero value, $c^{(N+1)}_{lm}=0.$ When making a matrix smaller, drop the unmappable, largest momentum coefficients. This procedure makes sense as the highest moments on the $(N+1)$ sphere are above the cut-off of the $(N)$ sphere.

Now, we use these two maps to define the first and second derivatives for a given layer as
\begin{equation}\label{partialN1}
    \partial^{(N)}_r \Phi^{(N)} = \frac{{\cal{D}}\Phi^{(N+1)} - {{\cal{U}}}\Phi^{(N-1)}  }{2\lambda},
\end{equation}
and 
\begin{equation}\label{partialN2}
       \partial^{2\,(N)}_r \Phi^{(N)} = \frac{{\cal{D}}\Phi^{(N+1)} -2\Phi^{(N)} + {{\cal{U}}}\phi^{(N-1)} }{\lambda^2},
\end{equation}
clearly motivated by the finite version of the expressions
\begin{align}
f'(x)=\lim_{\ep\to 0} \frac{f(x+\ep)-f(x-\ep)}{2\ep}\ ,\ f''(x)=\lim_{\ep\to 0} \frac{f(x+\ep)-2f(x)+f(x-\ep)}{\ep^2}.\label{der3}
\end{align}
An issue arises on the innermost and the outermost layer, where there is no next layer to compare with. We thus define $\mathcal U \Phi^{(M)}$ and $\mathcal D \Phi^{(1)}$ to vanish, which is consistent with zero Dirichlet boundary conditions at the outer layer\footnote{A different way to view this is that the value of any function anywhere outside the ball of radius $\lambda M$ is zero.} but is a new condition, put in by hand, at the inner layer. However, in the large $M$ limit, this should not play any role. In some examples, such as the heat transfer equation, one might prefer to choose the Neumann boundary condition instead. 

We can now define the radial part of the Laplace operator as
\begin{equation} \label{KR}
    \mathcal K_R \Psi= \partial^2_r\Psi+2 r^{-1}\partial_r\Psi\ ,\ \partial_r \Psi= \begin{pmatrix}
\partial^{(1)}_r \Phi^{(1)} &  &  &  &\\
 & \partial^{(2)}_r \Phi^{(2)} &  &  &\\
 &  & \partial^{(3)}_r \Phi^{(3)} &  &\\
 &  &  & \ddots  &\\
 &  &  &  & \partial^{(M)}_r \Phi^{(M)}
\end{pmatrix}\ ,
\end{equation}
and similarly for $\partial^2_r\Psi$. Here, we abused the notation a little since the action of $\partial_r$ is not truly block-diagonal and mixes values at different layers, i.e. different blocks. The action of $\partial_r$ can not be expressed as a simple matrix action of anything on the matrix $\Psi$; we will return to this issue in section \ref{sec5}. This can, in turn, be used to express the action of the radial part of the kinetic term as follows
\begin{equation} \label{KR2}
   {\cal{K}}_R \Psi = \sum \limits_{N,l,m}\frac{(N+1) c^{(N+1)}_{lm} + (N-1) c^{(N-1)}_{lm} - 2 N c^{(N)}_{lm}}{N\,\lambda^2} Y^{(N)}_{lm}.
\end{equation}

Clearly, the choice of the derivatives (\ref{partialN1},\ref{partialN2}) is to some extent ambiguous, as is the choice of the radial Laplacian in \eqref{KR}; with our choice $\partial_r\partial_r\neq \partial_r^2$. We could have as well used $r^{-2} \partial_r r^2 \partial_r \Psi$ there and let ourselves be motivated by different, perhaps more precise or more compatible, finite differences in \eqref{der3}. We have experimented with other choices and will comment on these attempts where appropriate. This choice is justified because it is simple and leads to reasonable results. Since we are using three layers to compute the second derivative, our results agree up to the second order with the continuum limit. The $f'''(r)$ correction can be removed by including two more layers in the calculation and so on for higher derivative corrections.
There are also boundary effects at the outermost layer due to the Dirichlet boundary condition. In principle, these could be alleviated by Neumann boundary conditions, which we do in the study of heat transfer.
Also, one can show that with this definition of the radial kinetic Laplacian, matrix $r^{-1}$ is the Green's function up to the above-mentioned boundary effect, i.e. $\K_R r^{-1}$ yields a function which is positive on the innermost layer and zero elsewhere and that that $\mbox{Tr}\lr{\K_R r^{-1}}=1$.

Together, we are finally set to define the fuzzy onion space ${\cal{O}}^3_\lambda$ as a configuration space of block diagonal matrices $\Psi$ of the form \eqref{psi} equipped with the Laplace operator that introduces the geometry of the space as governed by \eqref{KL} and \eqref{KR}. Together, we define the kinetic operator on the fuzzy onion as
\begin{equation}\label{K}
{\cal{K}} = {\cal{K}}_L + {\cal{K}}_R.
\end{equation}
This construction produces a 3-dimensional space of concentric fuzzy spheres of increasing radius. Note an interesting feature --- while the spherical quantization has a momentum-cut-off structure on each layer, the radial direction is quantized in a lattice-like way. The same structure was observed for $\mathbb R_\lambda^3$ space \cite{pgk15}. The most appealing feature of this case is that the model is formulated in terms of Hermitian matrices and thus is accessible for numerical simulations. 

\textcolor{black}{Let us briefly discuss how this formulation compares to some previous constructions. As we will see in section \ref{sec5:col}, our definition seems to be compatible with the Laplace operator \eqref{H0} from \cite{pgk15}. It differs from the Laplace operator of \cite{seik,patricia}, where the radial part only couples modes on the same layer and does not connect different layers together. It would be interesting to see how the construction of \cite{diffstar,todorovic2} compares to ours, but since it is not carried out in a matrix base, it is unclear what the connections are. The introduction of the matrix basis is not completely new and was introduced as early as \cite{patricia}. Our proposal, however, does this explicitly, which makes the structure of the functions and operators more transparent. It also allows for the use of well-established matrix methods to be presented in the next section.}

\section{Examples} \label{sec4}

With the definition of $\Psi$ and ${\cal{K}}$, one can do a lot of physics. We have chosen three examples. \textcolor{black}{These are meant to illustrate in different physically relevant and interesting situations the workings, advantages and limitations of our construction. This section is thus a proof-of-concept of the matrix formulation and the radial Laplacian \eqref{KR}.}

The first example is the MC study of the scalar field theory, one of the most thoroughly studied examples of the fuzzy sphere model. The second example is heat transfer, showing that one can do classical physics on a fuzzy space. The third example is the quantum mechanical Coulomb problem, as it was studied in the bosonic formulation of the three-dimensional noncommutative space that should be, in results, similar to the fuzzy onion model presented here.

\subsection{The $\Phi^4$ scalar field theory}

A scalar field theory can be defined in a straightforward way using the matrix action 
\begin{equation} \label{action}
    S[\Psi] = 4\pi \lambda^2 \mbox{Tr } r\left( a\ \Psi {\cal{K}} \Psi + b \ \Psi^2 + c\ \Psi^4 \right), \ \ {\cal{K}} =  {\cal{K}}_R + {\cal{K}}_L.
\end{equation}
The mean value of observables is defined in the usual way:
\begin{align}
    \left\langle {\cal {O}}(\Psi)\right\rangle=\frac{1}{Z} \int d\Psi e^{-S(\Psi)}{\cal {O}}(\Psi),\ d\Psi=\prod_{N=1}^{M}d\Phi^{(N)}\ .\label{expectation}
\end{align}
The integration goes over all matrices of the form \eqref{psi} and can be evaluated numerically using the Hamiltonian Monte Carlo method (HMC) in the same way as for other fuzzy spaces \cite{Ydri:2014rea,Rea:2015wta, Bosetti:2015lsa,Kovacik:2018thy}. There is an additional difficulty in computing the kinetic term; the simulation has to compute the Fourier transformation at every step. One can, in principle, set up the simulation primarily in terms of the expansion coefficients, but then defining the momentum matrix in HMC is suitably being done in $\Psi$-representation and one has to perform the Fourier transformation nonetheless. 

Let us briefly discuss the behaviour of the scalar field theory on fuzzy spaces in general. The theory is expected to recover the behaviour of its continuous counterpart in the infinite matrix size limit. With the action \eqref{action}, one can choose $a=1$. The continuous theory then has two phases depending on the value of $b$ and $c$. Above certain value, $b>b_{\scriptsize{c}}(c)$ the field has zero expectation value, below it becomes nonzero. In fuzzy spaces, the existence of a third phase has been observed where the field oscillates in a stripe-like fashion between two nonzero values; for a review see \cite{stripes}.

Usually, one characterizes matrix theory by the probabilistic distribution of the matrix eigenvalues. The striped phase mentioned above is characterized by an eigenvalue distribution defined on two separate intervals. This phase is also called the nonuniformly ordered phase. 

\begin{figure}

\centering
\begin{subfigure}[b]{0.90\textwidth}
   \includegraphics[width=1\linewidth]{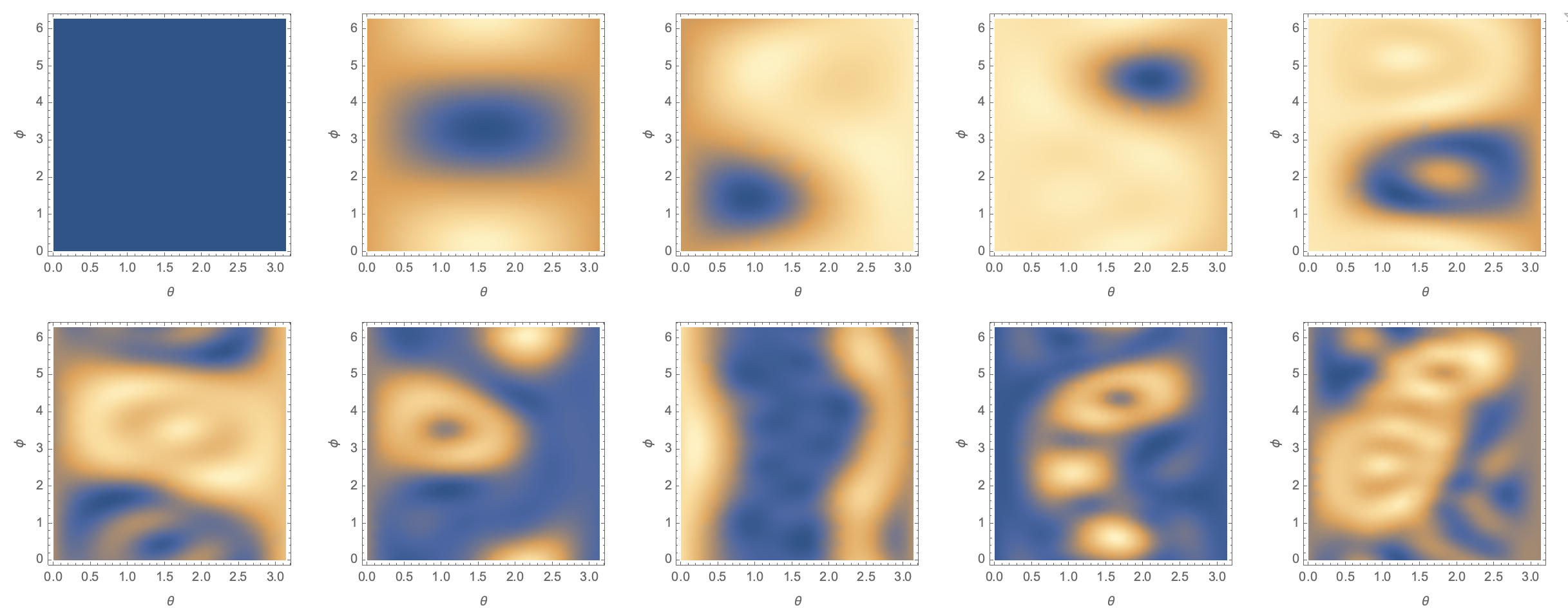}
   \caption{} 
\end{subfigure}

\begin{subfigure}[b]{0.90\textwidth}
   \includegraphics[width=1\linewidth]{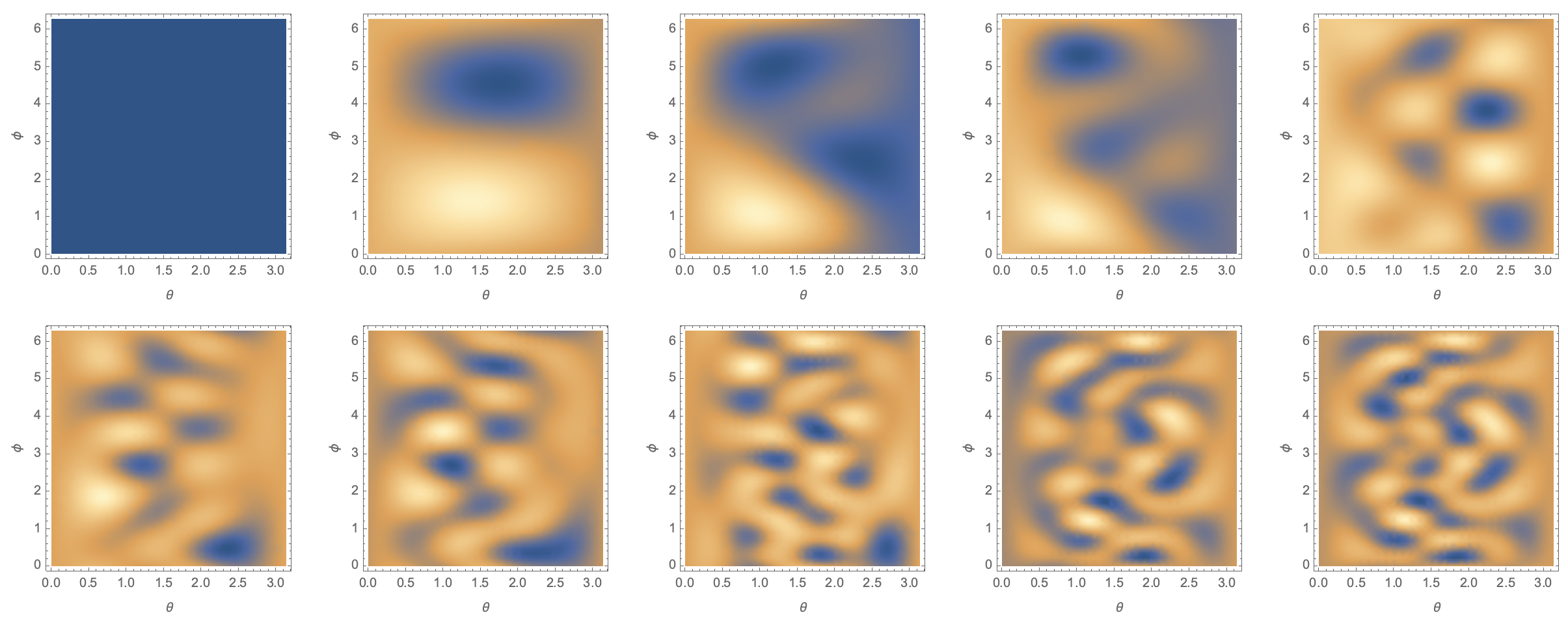}
   \caption{}
\end{subfigure}
\caption{Plots of $\Phi(\theta,\varphi)$ obtained using \eqref{map} for each of the layers from an HMC simulation using \eqref{action} with $a=1, b= - 3, c = 5$ and $N=10$ without the radial part of the kinetic term (panel a) and with it (panel b). We can observe that the fields are more correlated for given $\theta$ and $\varphi$ (not jumping between negative and positive values) when the radial part of the kinetic term is included.}

\label{fieldtheory}
\end{figure}

As an example, we have set up the model with the maximal matrix size $M=10$ and coefficients $a=1, b= -3, c = 5$, which is expected to be in the nonuniformly ordered phase. The simulations were initiated from a state of all elements being zero and thermalized, so one can expect the eigenvalues to split between two different minima. We have used the HMC algorithm as described in \cite{Kovacik:2022kfh} with the eigenvalue procedure flipped off as it has not been tested in this context before. The step-length in simulations has been tuned to reach around $80\%$ acceptance rate and checked the Schwinger constrain to be within $1\%$. For this same setup, we ran two simulations: one with the full kinetic term, \eqref{K} and one without the radial part, ${\cal K}_R$. After a sufficient number of steps, of the order of $10^6$, the current state $\Psi$ has been saved and using the map \eqref{map} translated into spherical harmonics $Y_{l,m}(\theta,\varphi)$ which are shown in the figure \ref{fieldtheory}.

In the case without the radial part of the kinetic term, the fields on different layers are separated --- as seen in the upper part of the figure \ref{fieldtheory}. The fluctuations on individual layers are not coupled in the radial direction. Also, notice that lower layers contain less detailed structures than the upper ones, as the smaller matrices have fewer degrees of freedom. On the other hand, the bottom part of the image shows the theory with the radial part of the kinetic term included. Notice that fields across various layers are organized, and for given values of $\theta$ and $\phi$, the fluctuations on different layers are similar.

\subsection{Heat transfer}

With the kinetic term \eqref{K}, we are set up to study many classical systems, such as wave propagation or heat transfer. We opt for the latter example, but it is straightforward to write down equations for the other cases. The heat transfer equation reads:
\begin{equation} \label{heatEQ}
    {\cal K} \Psi (t)= \alpha\ \partial_t \Psi(t),
\end{equation}
Here $\Psi(t)$ is understood as a time-dependent matrix of the form \eqref{psi}. In numerical simulations, we consider it as a sequence of matrices ${\Psi_i, i = 0,\ldots, T/\Delta t}$ and where ${t_i = t_0 + i \Delta t}$ is the discretized time. 

\begin{figure}
\includegraphics[width=1.\textwidth]{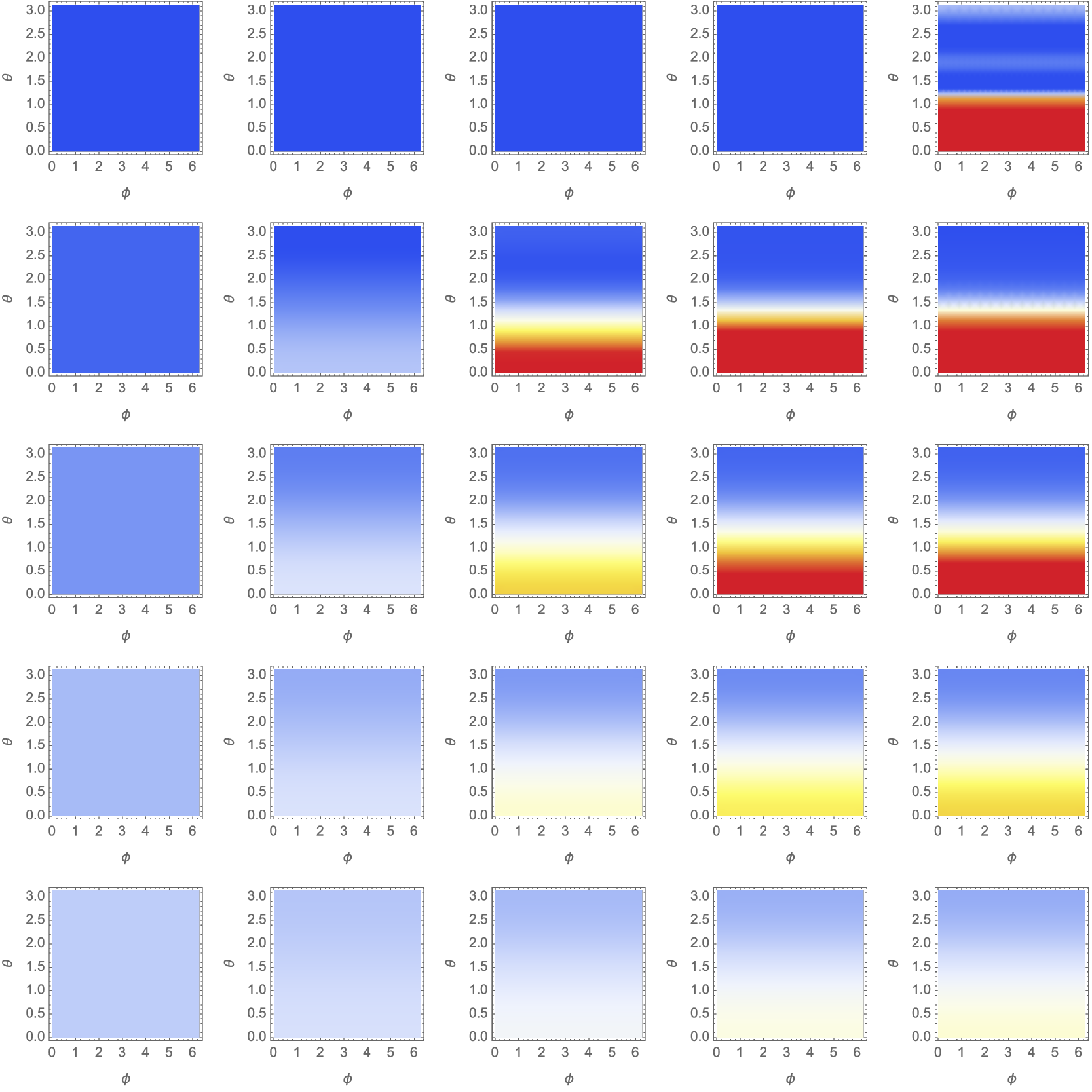} 

\caption{Simulation of the heat transfer from \eqref{heatEQ} with $\alpha = 1, M=5$. Horizontally, we have fields on five different layers; vertically, the time goes from top to bottom in steps $t_i=0, 0.2, 0.4, 0.6, 0.8$. It can be seen that the heat initially positioned close to a pole of the outermost layer slowly dissipates across all layers and moves toward thermal equilibrium.}
\label{heat}
\end{figure}

As an example, we have initiated the model with ${M=5}$ in a state with a single nonzero element at the outermost layer:
\[\Phi^{(M)}=
\begin{pmatrix}
1 & 0 & \\
0 & 0 & \\
  &   &\ddots 
\end{pmatrix}
\mbox{ and } \Phi^{(M')}=
\begin{pmatrix}
0 & 0 & \\
0 & 0 & \\
  &   &\ddots 
\end{pmatrix} \mbox{ for } M'<M.\]
This means that the temperature was initially zero everywhere but near the north pole of the outermost layer. We have used the Neumann boundary condition so there was no leakage from the innermost and outermost layers. The time evolution was obtained by Euler integration, with the time step small enough to avoid numerical instabilities. We can see that with $\alpha = 1$ after a total time of $0.8$ has passed, the heat has been distributed nearly evenly across all layers; see figure \ref{heat}.

In principle, it is possible to search for the eigenstates of the kinetic term, that is, solutions of the equation:
\begin{equation} \label{heatEQeigen}
    {\cal K} \Psi_n = \lambda_n \Psi_n, \mbox{ where } \lambda_n \le 0\ ,
\end{equation}
and then expanding the initial state into modes $\Psi_n$. All of them but the one with $\lambda_0=0$ decay exponentially. Nonetheless, we have opted for the numerical simulation to test our code. Also, numerical methods make adding new features, such as time-dependent heat sinks or sources easy.

\subsection{The Coulomb problem}\label{sec5:col}

Let us now investigate the problem of Schr\"odinger equation with Coulomb potential in the fuzzy onion space. As mentioned above, this quantum-mechanical situation has been considered in a different formalism \cite{pg13,pgk15}, so comparing those results with our formulation offers important insight. The Hamiltonian is given by
\begin{align}\label{HatomH}
    H=-\frac{\hbar^2}{2m_e}\K-q r^{-1}
\end{align}
where we taken ${\hbar=m_e=q=1}$ and look for solutions of the eigenvalue problem\footnote{This also sets the Bohr radius ${a_0=1}$.}
\begin{align}
    H \Psi =E \Psi\ .
\end{align}
Recall that the Coulomb problem in ordinary quantum mechanics is solved by splitting the solution between the radial and angular parts. We can do something similar by fixing $(l,m)$ and solving only for the radial part. Notice that one can think of the matrix $\Psi$ we have been using so far as an array of matrices $\Psi = \left( \Phi^{(1)}, \Phi^{(2)},\ldots\right)$. However, when $(l,m)$ is fixed, each layer is represented only by the single coefficient $c^{(N)}_{lm}$ and the entire function is represented by 
\begin{eqnarray}
 {\cal C}_{lm} = \left(c^{(l+1)}_{lm} ,\ldots,c^{(M)}_{lm}\right).  
\end{eqnarray}
Note that only spheres with $N>l+1$ can carry states with the angular momentum $l$ so those below have been omitted without losing any information. Alternatively, we can define them to be zero. 
In the next section, we will reformulate in detail the model in terms of the vector $\C$ and rewrite operators as ${d\times d}$ matrices acting on such vectors. For the moment, let us denote representation in terms of such matrices by boldface letters and continue. 

We have the Schr\"odinger equation of the form
\begin{align} \label{Schrodinger}
    \mathbf{H} {\cal C}_{lm} =E {\cal C}_{lm}\ 
\end{align}
Since the Hamiltonian and the space are radially symmetric, we expect the energies not to depend on $l$ and $m$ and for the sake of simplicity, we can analyze the $l=0$ state. So the $\cal C$ is an $M$-dimensional vector and the Hamiltonian is
\begin{align}
    \mathbf H=-\frac{1}{2}\mathbf K _R-\mathbf r^{-1}\label{num:ham1}
\end{align}
The upshot of the $l=0$ restriction is also the fact that the matrix representation of the derivative operators (\ref{kin1},\ref{partialmat},\ref{partialmat2}) is a straightforward matrix with nonzero entries on the main diagonal and/or the next two diagonals above and below. The matrix-eigenvalue problem \eqref{Schrodinger} can be solved by usual methods --- for example numerically, as we have done for $M$ up to 3200.\footnote{We have also computed the spectrum of a matrix corresponding to nonzero values of $l$ and $m$, for smaller $M$ however. The results confirm that the energies indeed do not depend on $l$ and $m$ as expected.}

\subsubsection*{Comparison of the results}

By first choosing ${M=50}$ and ${\lambda=1}$, we can calculate the eigenvalues of matrix $\mathbf{H}$ to obtain the values given in table \ref{tab1}. We obtained 6 negative eigenvalues, naively representing bound states. These values are greater than the standard values in continuous spaces, in accordance with the intuition of the wave function of the electron being smeared away from the origin due to the nonlocality of the space and being squeezed into a volume region of the fuzzy onion. However, the most striking feature is how well these results reproduce the energies \eqref{pgk:energiesI} obtained previously in \cite{pg13,pgk15}. The ground state energy is numerically consistent in the first 35 digits, the second in the first 15 and only the last one differs considerably. The situation is even better when we increase the size of the matrix, and for ${M=300}$ we obtain 15 negative energy levels, the first two in more than 100-digit agreement with \eqref{pgk:energiesI} and the rest also reproducing this formula with striking accuracy.

The other thing that we need to look at is the behaviour of the energy levels by changing $\lambda$. Table \ref{tab2} gives these values for various pairs of $\lambda$ and $M$. It can be summarized as follows. As we lower $\lambda$ for a fixed number of layers, the energy levels start to differ significantly from \eqref{pgk:energiesI} and stop resembling the hydrogen atom problem completely. This is, however, to be expected because we shrink the region where the fuzzy structure exists, and the space becomes dominated by the outside region, where no dynamics at all are defined. But we can also see that we can always increase the number of layers such that the spectrum again becomes very well described by \eqref{pgk:energiesI}. Since this formula reproduces the spectrum of standard commutative quantum mechanics, it is reasonable to expect that the simultaneous limit ${\lambda\to0, M\to\infty}$ of the fuzzy onion as defined in section \ref{sec3} does so too. Keep however in mind that in order to recover the full three-dimensional commutative space, one must take the limit in such a way that $R = \lambda M \rightarrow \infty$ (in theory) or $\lambda M \gg a_0$ (in practice), where $a_0$ is the Bohr radius.

\begin{table}[h!]
    \centering
    \begin{tabular}{c||c|c|c|c|c|c}
        $n$ &  1 & 2 & 3 & 4 & 5 & 6\\
        \hline
        $E_n$ & -0.4142 & -0.1180 & -0.0541 & -0.0307 & -0.0179 & -0.0031\\
        \hline
        $E^{I}_{\lambda n}$ & -0.4142 & -0.1180 & -0.0541 & -0.0307 & -0.0198 & -0.0138\\
        \hline
        $E^{CQM}_{n}$ & -0.5 & -0.125 & -0.0556 & -0.0313 & -0.02 & -0.0139
    \end{tabular}
    \caption{Negative eigenvalues of the Hamiltonian \eqref{num:ham1} for ${M=50}$ and ${\lambda=1}$, with the values \eqref{pgk:energiesI} and appropriate energies of the hydrogen atom in the standard quantum mechanics.}
    \label{tab1}
\end{table}

\begin{table}[]

    \begin{tabular}{c||c|c|c}
        $M $ & $\lambda=0.1$ & $\lambda=0.01$ & $\lambda=0.001$ \\
        \hline
        $50$ & $6.24\cdot 10^{-3}$ & N/A & N/A \\
        \hline
        $100$ & $1.27\cdot 10^{-6}$ & N/A & N/A \\
        \hline
        $200$ & $1.97\cdot 10^{-13}$ & $2.81$ & N/A  \\
        \hline
        $400$ & $1.56\cdot 10^{-13}$ & $3.41\cdot 10^{-2}$ & N/A \\
        \hline
        $800$ & $5.22\cdot 10^{-13}$ & $4.9\cdot 10^{-5}$ & N/A \\
        \hline
        $1600$ & $4.8\cdot 10^{-14}$ & $1.13\cdot 10^{-11}$ & N/A\\
        \hline
        $3200$ & $9.02\cdot 10^{-15}$ & $5.75\cdot 10^{-12}$ & $1.26\cdot 10^{-1}$
    \end{tabular}    
    \begin{tabular}{c||c|c|c}
        $M $ & $\lambda=0.1$ & $\lambda=0.01$ & $\lambda=0.001$ \\
        \hline
        $50$ & N/A & N/A & N/A \\
        \hline
        $100$ & $1.01\cdot 10^{-1}$ & N/A & N/A \\
        \hline
        $200$ & $9.61\cdot 10^{-5}$ & N/A & N/A  \\
        \hline
        $400$ & $6.05\cdot 10^{-12}$ & N/A & N/A \\
        \hline
        $800$ & $6.72\cdot 10^{-12}$ & $4.7\cdot 10^{-1}$ & N/A \\
        \hline
        $1600$ & $1.53\cdot 10^{-12}$ & $1.99\cdot 10^{-3}$ & N/A\\
        \hline
        $3200$ & $9.99\cdot 10^{-13}$ & $4.97\cdot 10^{-9}$ & N/A
    \end{tabular}
    
    \caption{Relative difference of the eigenvalues of the Hamiltonian \eqref{num:ham1} and the corresponding values from \eqref{pgk:energiesI} for different values of $M$ and $\lambda$. The tables are for the $n=1$ and $n=2$ eigenvalues, respectively. N/A means that the given Hamiltonian does not have enough negative eigenvalues. One can see excellent agreement in the cases where the condition $M \lambda \gg 1$ is satisfied.}
    \label{tab2}
\end{table}

This is not a rigorous proof that Laplacians \eqref{K} (or \eqref{kin1}) and \eqref{H0} are equivalent, but strongly indicates that this is the case, at least for our needs. We thus conclude that, presumably, the radial Laplacian \eqref{K} reproduces the dynamics of the Laplacian \eqref{H0} considered in \cite{pgk15}, including the correct commutative limit. We leave rigorous proof of this statement for future work.

It would be interesting to see whether the spectrum of Hamiltonian \eqref{kin1} also includes the scattering states, or at least states that become the scattering states in the large $M$ limit; the found positive energy states are natural candidates. Another unanswered question is whether the spectrum includes the positive energy bound states in \eqref{pgk:energiesI}. We leave these questions for future work, too.

\begin{figure}
\includegraphics[width=0.48\textwidth]{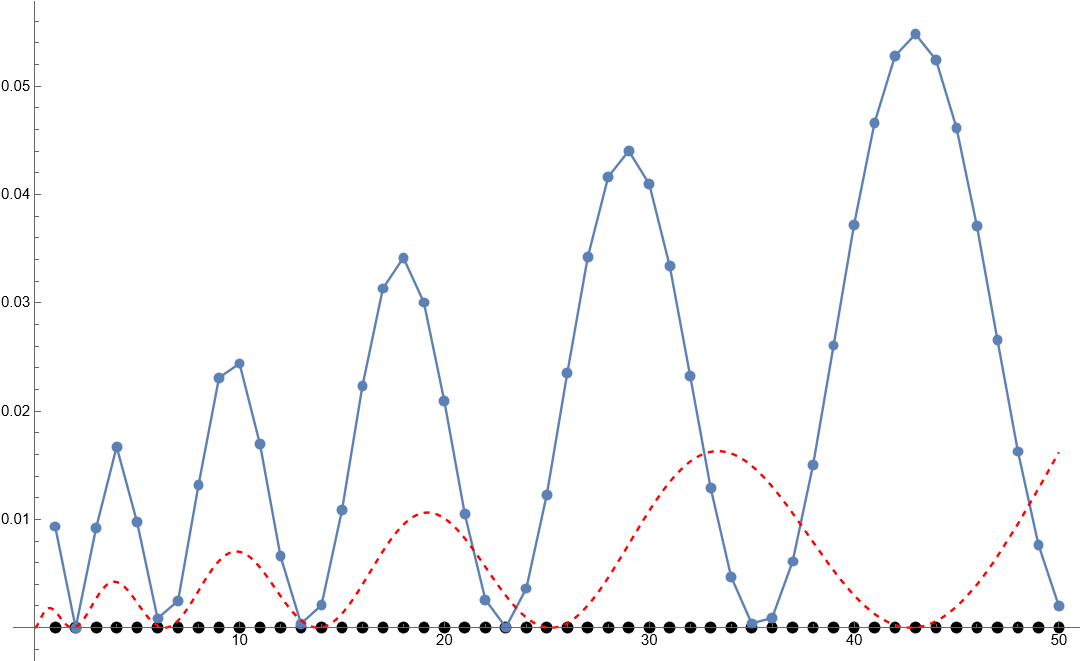} 
\includegraphics[width=0.48\textwidth]{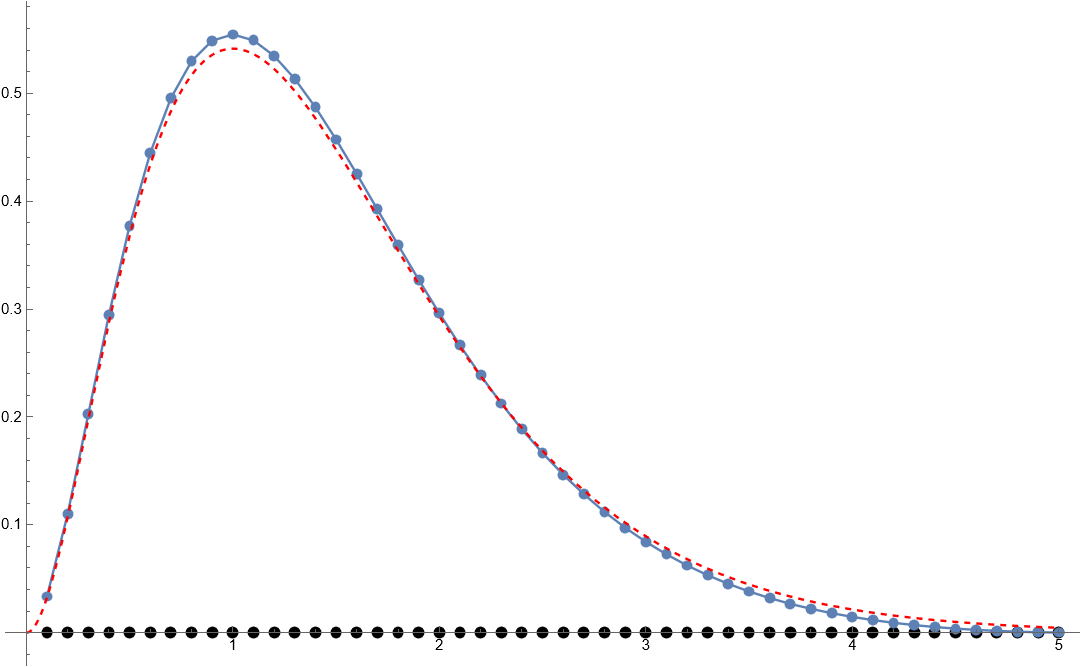} 
\includegraphics[width=0.48\textwidth]{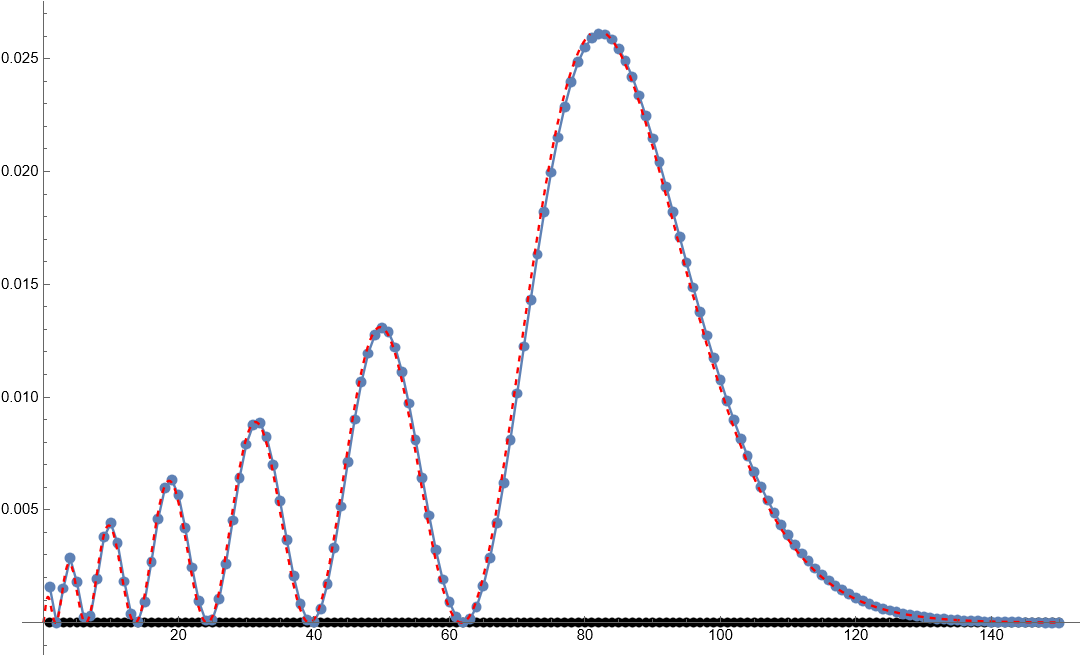} 
\includegraphics[width=0.48\textwidth]{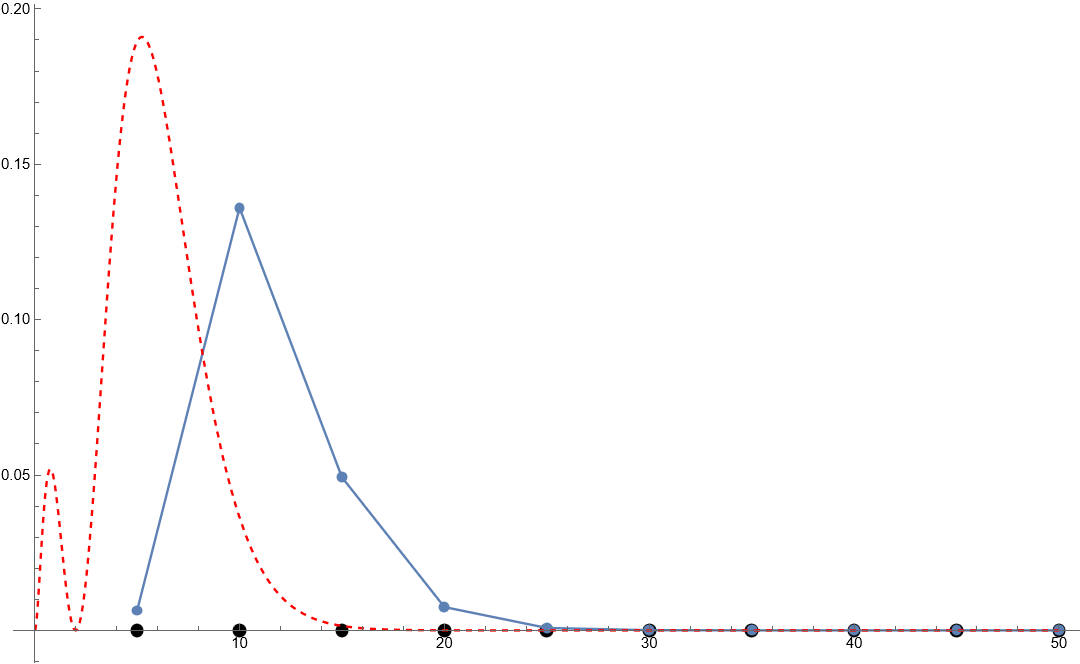} 
\includegraphics[width=0.48\textwidth]{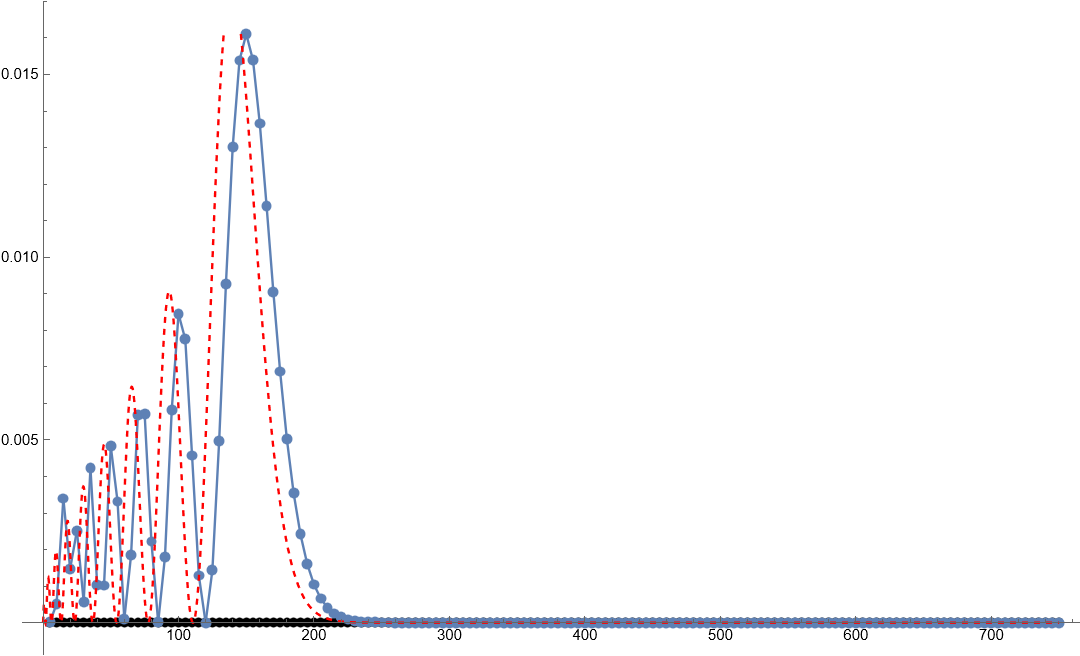} 
\includegraphics[width=0.48\textwidth]{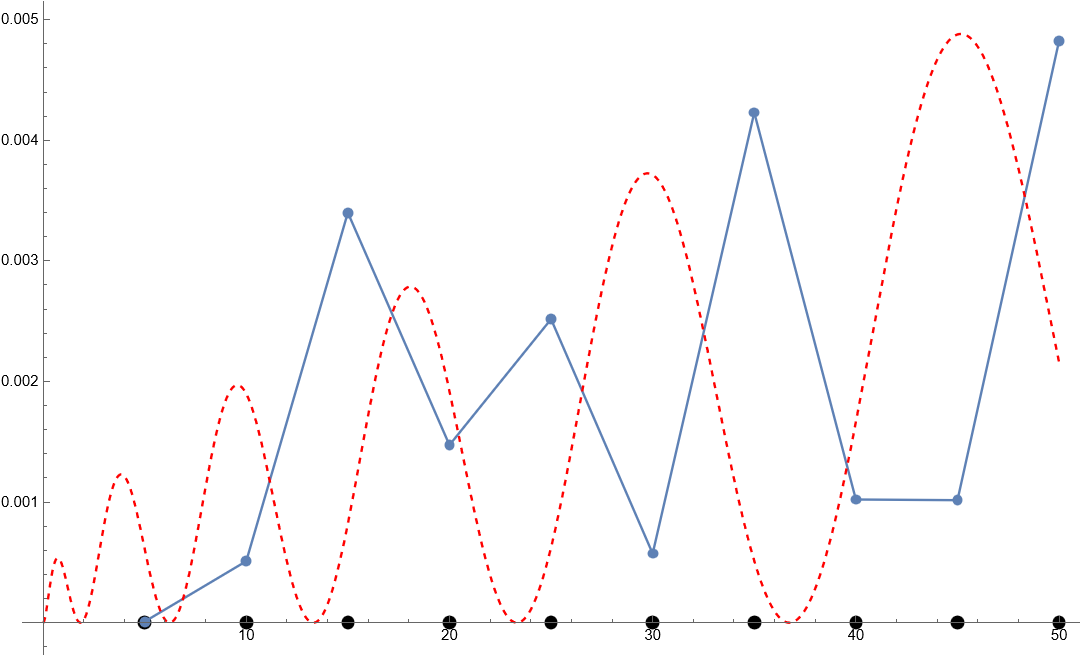} 
\caption{In all the figures, the horizontal axis gives the distance from the origin in units of the Bohr radius, which equals $1$ in our units. The black dots denote the location of the onion layers in the particular case. The blue line represents the radial probability distribution for the electron in the fuzzy onion space $\mathbb R_\lambda^3$ for the given values of parameters. The dashed red line represents the radial probability distribution of the corresponding state in the continuous space $\mathbb R^3$. The values of the parameters are, left to right, top to bottom, as follows:
 ${M=50,\lambda=1,n=6}$;
 ${M=50,\lambda=0.1,n=1}$;
 ${M=150,\lambda=1,n=7}$;
 ${M=50,\lambda=5,n=2}$;
 ${M=150,\lambda=5,n=9}$;
 ${M=150,\lambda=5,n=9}$ again, but a closeup to the origin. For discussion of the plots, see the main text.}
\label{fig:states}
\end{figure}

We can go beyond the energy levels of the hydrogen atom and look at the corresponding electron distributions. These were obtained from the eigenvectors of \eqref{Schrodinger}. We show several examples of radial probability distributions in the figure \ref{fig:states}. This will help us gain insight into what is happening in the system. If the fuzzy onion is not large enough, i.e. the classical electron has a significant part of its wave function outside, the NC wave function gets squeezed, considerably increasing its energy. If the classical wave function is almost completely localized within the fuzzy onion, we see two different behaviours. Suppose the length-scale $\lambda$ is comparable to or smaller than the Bohr radius. In that case, the radial probability for the electron has essentially the same features as in the classical case, being slightly repelled away from the origin. If, however, $\lambda$ is greater than the Bohr radius, there are features of the classical wave function that fall completely within the one step in the radial direction and are washed away in the noncommutative case. And the noncommutative wave function starts to resemble the classical distribution only when the features extend over distances larger than $\lambda$. All of this behaviour was to be expected and confirms our intuition about the model.

To conclude this section, let us briefly mention the results for different Laplacians. We have obtained the energy levels for a symmetrized version of \eqref{KR} (including a finite lattice-like correction) and results for a matrix version of the ${r^{-2}\partial_r r^2 \partial_r}$ case. In the first case, we obtained a spectrum that was in complete contradiction with the expected commutative limit and with the spectrum \eqref{pgk:energiesI} of \cite{pg13,pgk15}. In the latter case, however, we have obtained a reasonable spectrum with energy doublers and energy levels that were different from \eqref{pgk:energiesI}.

\section{The fuzzy onion as a vector model}\label{sec5}

\subsection{Definition of the model}
The inconvenient --- but perhaps necessary --- part of our construction of the fuzzy onion model is the necessity to keep computing the expansion coefficients $c$, while working in the matrix base $\Phi^{(N)}$. We will show here that the model can be expressed fully in terms of the expansion coefficients. In practical cases, such as examples analysed before, this comes with additional difficulties but we still find it useful, at least to understand the model better.

First, let us rephrase the functions on a single layer $\Phi^{(N)}$ as a vector model of dimension $N^2$. The advantage of such a formulation is that the Laplacian term \eqref{K} will now be expressed as an action of a single matrix. We first expand the matrix $\Phi^{(N)}$ into a hermitian basis\footnote{For simplicity, we will denote the pair of indexes $l,m$ by one index $\mu$, such that $l=m=0$ corresponds to ${\mu=0}$.} $T_\mu$
\begin{align}
\Phi^{(N)}=\sum_{\mu=0}^{N^2-1}c^{(N)}_\mu T^{(N)}_\mu=c^{(N)}_0T^{(N)}_0+\sum_{a=1}^{N^2-1}c^{(N)}_a T^{(N)}_a\ ,
\end{align}
which is normalized so that
\begin{align}
    \mbox{tr}_N \lr{T^{(N)}_\mu T^{(N)}_\nu}=\half\delta_{\mu\nu}\ .
\end{align}
See appendix \ref{appA} for our other conventions and some usefull identities. We will denote the $N^2$-dimensional column vector of $c$'s as
\begin{align}
    \C^{(N)}=\lr{c^{(N)}_0,c^{(N)}_1,\ldots,c^{(N)}_{N^2-1}}^T\ .
\end{align}
In the hermitian basis, the $c$'s of a hermitian matrix $\phi$ are all real and independent. The action of the kinetic term is given by
\begin{align}
    \mathbf{K}^{(N)} T^{(N)}_a= l(l+1)T^{(N)}_a\ ,\label{Teigen}
\end{align}
with $l$ being the corresponding angular momentum, we refrain from using the subscript to have a less cluttered notation. In what follows, we will do our best to avoid confusion by denoting operators in their ${N^2\times N^2}$, or later ${d\times d}$, matrix version by boldface letters. There are $N$ different possible values for $l =0,1,\ldots,N-1$.

For the full fuzzy onion, we can now write
\begin{align}\label{Cs}
    \C=\lr{\C^{(1)},\C^{(2)},\ldots,\C^{(M)}}^T\ ,
\end{align}
which is a $d$-dimensional \eqref{dimd} vector and essentially gives the coefficients of the matrix $\Psi$ in the basis
\begin{align}
T_A=\begin{pmatrix}
\mathbb{0}_{1\times 1} &  &  &  &\\
& \ddots  &  &  &\\
&  & T_a^{(N)} &  &\\
&  &  & \ddots  &\\
&  &  &  & \mathbb{0}_{M\times M}
\end{pmatrix}\ , \ A=1,\ldots,d\ .
\end{align}

Let us now define other operators as acting on $\C$ in \eqref{Cs}. The matrix $r$ is a block diagonal matrix
\begin{align}\label{rmat}
    \mathbf{r}=\begin{pmatrix}
\lambda \one_{1\times 1}&  &  & &\\
 & 2 \lambda \one_{4\times 4}  &  &  &\\
 &  & 3 \lambda \one_{9\times 9} &  &\\
 &  &  & \ddots  &\\
 &  &  & & M \lambda \one_{M^2\times M^2}
\end{pmatrix}
\end{align}
of matrices $ \lambda N \one_{N^2\times N^2}$ for $N=1,2,\ldots,M$. Note that before $r$ in \eqref{r} has been a similar block diagonal matrix, however, with the size of blocks $N$. This is a crucial difference. Before, the matrix $r$ acted on matrix $\Psi$ by matrix multiplication; now, it acts on the vector $\C$. Before, there was no nice way to express the action of the derivatives $\partial_r$ on the function $\Psi$; now, they act as matrices on vector $\C$. The angular part of the kinetic term is block diagonal
\begin{align}\label{KLmat}
    \mathbf{K}_L=\mathbf{r}^{-2}\begin{pmatrix}
\mathbf{K}^{(1)} &  &  & &\\
 & \mathbf{K}^{(2)}  &  & &\\
 &  & \mathbf{K}^{(3)} & &\\
 &  &  & \ddots &\\
  &  &  & & \mathbf{K}^{(M)}
\end{pmatrix}\ ,
\end{align}
with $\mathbf{K}^{(N)}$ being angular Laplacian on the given layer, itself a block diagonal matrix of matrices $l(l+1)\one_{(2l+1)\times(2l+1)}$ for $l=0,1,\ldots,N-1$ due to \eqref{Teigen}.

The radial part \eqref{KR}, when interpreted as a matrix acting on the vector $\C$, is off-diagonal even in the block sense as it connects terms across various layers. Since operators $\mathcal U$ and $\mathcal D$ used to express the derivatives \eqref{partialN2} are defined in terms of the coefficients $c^{(N)}_{lm}$, it is straightforward to write the $d\times d$ version $\mathbf{D}_R$ of $\partial_r$ and $\mathbf{D}^2_R$ of $\partial_r^2$ from \eqref{partialN2}, with explicit formulas given in the appendix \ref{appD}.

Finally, let us repeat the definition of the Laplacian, now understood as a ${d\times d}$ matrix acting on a $d$-dimensional vector space of $\C$'s:
\begin{align}\label{Kmat}
    \mathbf{K}\,=\,&\mathbf{K}_L+\mathbf{K}_R\ ,\\
    \mathbf{K}_R\,=\,&\mathbf{D}^2_R+2 \mathbf{r}^{-1} \mathbf{D}_R\ ,\label{kin1}
\end{align}
with matrices $\mathbf{r},\mathbf{K}_L,\mathbf{D}_R,\mathbf{D}^2_R$ given by (\ref{rmat},\ref{KLmat},\ref{partialmat},\ref{partialmat2}).

Once again, let us stress that this formulation is useful because the kinetic term acts explicitly as a matrix here. The drawback is that the function multiplication, before a straightforward matrix multiplication, is now rather complicated. In \cite{patricia}, a spectrum of the Laplacian constructed in \cite{seik} was calculated on the same basis, and it was shown that the Laplacian is block-diagonal, i.e. it does not couple modes on different layers of the onion. This is different from the above construction, where $\mathbf K_R$ part of the Laplace operator is nonvanishing also in the off-diagonal blocks.

\subsection{Field theory on the fuzzy onion as a random vector model}

As we have seen, the functions on the fuzzy onion can be thought of in terms of $d$-dimensional vectors and the operators acting on them in terms of ${d\times d}$ matrices. In this section, we would like to investigate the consequences of such an approach for fuzzy field theory. First, let us look at how this translates for functions on a single layer.

As mentioned in section \ref{sec3}, the action for the field theory is
\begin{equation}
S_N [\Phi^{(N)}] = \frac{4\pi}{N}\mbox{tr}_N \left( a\ \Phi^{(N)} {\mathcal{K}}^{(N)} \Phi^{(N)} + b\ (\Phi^{(N)})^2 + c\ (\Phi^{(N)})^4 \right),
\end{equation}
The quadratic part of this expression is straightforward, and we obtain
\begin{align}
    S_{N,0}=\frac 1 2\ \lr{\C^{(N)}}^{T} \cdot \mathbf{P} ^{-1} \cdot \C^{(N)}\ \ ,\ \ 
    \mathbf{P} = \frac{N}{4\pi}\lr{2a\mathbf{K}^{(N)}+ 2b\,\mathbb{1}_{N^2\times N^2}}^{-1}\ ,
\end{align}
From now on, we will always assume the first vector in the expressions like above to be a transposed row and drop the explicit $T$. Since matrix $P$ is diagonal, its inverse is straightforward to compute.

The interaction term is more involved, but after some algebra, summarized in the appendix \ref{appB}, we obtain
\begin{align}
    \mbox{tr}_N\lr{\lr{\Phi^{(N)}}^4}=\frac{1}{4N}\lr{\C^{(N)}\cdot\C^{(N)}}^2+\frac{1}{8}\lr{\C^{(N)} \cdot \mathbf{G}^{(N)}_a\cdot\C^{(N)}}^2\ ,
\end{align}
where the interaction matrices are given by
\begin{align}
\mathbf{G}^{(N)}_a=\lr{\begin{array}{cc}
     0 & (v^{(N)}_a)^T \\
     v^{(N)}_a & D^{(N)}_a
\end{array}}\ ,\ \lr{D^{(N)}_a}_{ij}=2\trl{\left\{T^{(N)}_i,T^{(N)}_j\right\}T^{(N)}_a}\ ,\  (v^{(N)}_a)_b=\sqrt{\frac{2}{N}}\delta_{ab}
\end{align}
This leaves us with the vector model action
\begin{align}
    S^{(N)}=\frac 1 2 \C^{(N)} \cdot \mathbf{P}^{-1} \cdot \C^{(N)}+\frac{4\pi}{N}c\slr{\frac{1}{4N}\lr{\C^{(N)}\cdot\C^{(N)}}^2+\frac{1}{8}\lr{\C^{(N)} \cdot \mathbf{G}^{(N)}_a\cdot\C^{(N)}}^2}\ .\label{layerCaction}
\end{align}

When we now couple the layers together, we need to add the radial part of the Laplacian into the action and obtain
\begin{align}\label{Caction}
    S\,=\,&4\pi\mbox{Tr}\ r\left(a\Psi \K \Psi+b \,\Psi^2 + c\,\Psi^4 \right)\,=\,\nonumber\\ \,=\,& \half \C\cdot \mathbf{P}^{-1} \cdot \C+4\pi\lambda^3\sum_{N=1}^{M}cN\slr{\frac{1}{4N}\lr{\C^{(N)}\cdot\C^{(N)}}^2+\frac{1}{8}\lr{\C^{(N)}\cdot \mathbf{G}_a^{(N)}\cdot\C^{(N)}}^2}\ ,\\
    &\mathbf{P}\,=\,\frac{1}{4\pi\lambda^2}\lr{2a\mathbf{r}\mathbf{K}+2b\mathbf{r}}^{-1}
\end{align}
We can see that the propagator is now not diagonal. Moreover, the interaction part of this expression cannot be nicely expressed in terms of the vector $\C$. We would instead like to write something like
\begin{align}\label{suggestion}
    S=\half \C\cdot \mathbf{P}^{-1} \cdot \C+4\pi c \lambda^3\slr{\frac{1}{4}(\C\cdot\C)^2+\frac{1}{8}(\C\cdot \mathbf{G}_A\cdot\C)^2}
\end{align}
for the action, where $\mathbf{G}_A$'s is a set of ${d-M}$ block diagonal matrices with one of the $NG^{(N)}_a$'s in the proper place on the diagonal and zeros everywhere else. This would be a very different model from \eqref{Caction} and would introduce nonlocality of the interaction in the radial direction. In this case, the interaction is introduced among all the layers in the radial direction, which is perhaps more nonlocal than necessary but is still a reasonable first proposition.

Models such as (\ref{layerCaction},\ref{Caction},\ref{suggestion}) are random vector models which in principle, can be studied by corresponding analytical and numerical techniques. The first one is quite well understood as a matrix model, so it could be used as a sandbox for understanding the tools in this setting, and then the other two could be studied to obtain new results.
 
\section{Conclusion and discussion}

In this paper, we have introduced an explicit matrix formulation of three-dimensional fuzzy space with rotational symmetry called the fuzzy onion, ${\cal O}_\lambda$. The main idea was to connect matrices describing individual fuzzy spheres into one large matrix and use the standard angular kinetic operator for each layer. Neighbouring layers are connected by a radial derivative term computed in the base of expansion coefficients while disregarding those that cannot be matched due to different degrees of freedom. We have also shown how to formulate the model purely in terms of expansion coefficients, $\C$, which is illuminating, but we used the matrix formulation for practical purposes. 

This way, we have constructed a discrete 3-dimensional structure with two different behaviours. Angular discreteness is of a noncommutative nature with the full rotational symmetry. However, the radial direction is lattice-like and rigid, with finite steps between the layers. It would be interesting to see if one could alleviate this rigidness into something more fuzzy-like, e.g. by smearing the radial part of the function over several layers \cite{corfu22}.

We have investigated three different physical models on the fuzzy onion. For the ${\Phi^4}$ scalar field theory, without the radial part of the kinetic term, the structures appearing on each layer were disorganized. But they align when the radial part of the kinetic term is turned on, showing that the construction of the radial Laplacian indeed brings the layers into contact. We plan to analyze the phase structure of the theory in future research. We have shown that the model can be used in the classical setting for perhaps the simplest case --- heat transfer without a source --- and the spreading of the heat across layers facilitated by the radial Laplacian behaved in an expected way. Formulating something more demanding, such as Navier-Stokes theory, is left for future research. Finally, we have investigated the quantum mechanical hydrogen atom problem. In our case, the problem turned into finding eigensystems of large matrices, which a computer can do reasonably quickly. The agreement with the previous construction was beyond any expectations, and we can conclude that the results are equivalent after taking the radius of the onion space to infinity. This is, however, still left to be proved rigorously and working in the ${l=0}$ regime might help since the matrices for radial derivatives simplify significantly. Working with a finite size of $M$ yields interesting questions; for example, the Hamiltonian has a finite number of positive and negative eigenvalues. What dictates their ratio, and how do they match the spectrum known from previous studies? As before, we leave this for future work.

We have focused on building the model and analyzed physical systems mostly as a proof of concept. In future studies, we plan to investigate those in greater detail. As a three-dimensional space model, the fuzzy onion is a good place to test the phenomenological consequences of such a structure, e.g. for light propagation, the behaviour of matter or the dynamics of the space(time) itself.

In \cite{hsjt22} the field theory on the fuzzy sphere has been described on a different basis, formed by extended string-like objects called string states. The string's energy is given, in the large-$N$ limit, by the length of the string. In the model we have presented here, only the modes with ends at the same layer are present, and it does not contain the modes extending from one layer to a different one\footnote{We thank Harold Steinacker for this observation.}. One possible way to extend our construction would be to include these modes extending between the layers, together with the natural Laplacian, which is, however, going to be more complicated due to the finite $N$ effects. In \cite{diffstar} a star-product has been constructed, different from \cite{seik}, to define a noncommutative version of $\mathbb{R}^d$. It would be interesting to see the relationship of this construction's ${d=3}$ case to the one presented here.

One is also tempted to interpret the radial direction of the model as a temporal and not spatial coordinate. In this case, the model would describe an expanding quantum sphere --- perhaps a helpful toy model to study the expansion of quantum space with a growing number of degrees of freedom, allowing us to study the quantum origin of primordial fluctuations in the universe. Also, one can study relativistic objects with rotational symmetry, such as the Schwarzschild black hole in a quantum space \cite{Schupp:2009pt}. Here, the separation of layers $\lambda$ can be taken to depend on the radius $\lambda(r)$, or perhaps to be even made angular dependent. In this way, we would describe a space-time with a curved and deformed structure in a way that resembles a realistic onion even more.

\paragraph{Acknowledgments.}
This research was supported by 
VEGA 1/0025/23 grant \emph{Matrix models and quantum gravity} and MUNI Award for Science and Humanities funded by the Grant Agency of Masaryk University.

We thank the organizers of \emph{Gravity, Noncommutative Geometry, Cosmology} held at CMO-BIRS, Oaxaca, \emph{Large-N Matrix Models and Emergent Geometry} at ESI Vienna, Austria and \emph{Workshop on Noncommutative and Generalized Geometry in String theory, Gauge theory and Related Physical Models} held in Corfu, Greece, for the possibility of disseminating this research and to all the participants of these events for many useful comments and suggestions.

\appendix

\section{Conventions}\label{appA}

This section briefly overviews our $su(2)$ generator conventions and some important formulas. In this and the following section, we will deal only with quantities defined on a single fuzzy sphere with a fixed $N$, so we will drop the superscript $(N)$ distinguishing between the layers. We express the field on the basis of polarization tensors $T_\mu$ as follows
\begin{align}
\Phi=\sum_{\mu=0}^{N^2-1}c_\mu T_\mu=c_0T_0+\sum_{a=1}^{N^2-1}c_a T_a
\end{align}
and such that
\begin{align}
\strl{T_\mu}=&\sqrt{\frac{N}{2}} \delta_{0\mu}\ ,\ \textrm{or } T_0=\frac{1}{\sqrt{2N}}\mathbb{1}_{N^2\times N^2} \textrm{ and } \strl{T_a}=0\ ,\\
\strl{T_a T_b}=&\half \delta_{ab}\ ,\\
\strl{T_a T_b T_c}=&\frac 1 4 d_{abc}+\textrm{ \textit{antisymmetric}}\ ,\\
\strl{T_a T_b T_c T_d}=&\frac{1}{4N}\lr{\delta_{ab}\delta_{cd}-\delta_{ac}\delta_{bd}+\delta_{ad}\delta_{bc}}
+\frac 1 8 \lr{d_{abe}d_{cde}-d_{ace}d_{bde}+d_{ade}d_{bce}}\nonumber\\&+\textrm{ \textit{antisymmetric}}\ ,
\end{align}
where \textit{antisymmetric} stands for terms that are antisymmetric in a pair of indexes and thus will not be relevant in our calculations. Also
\begin{align}
    d_{abc}=2\strl{\left\{T_a,T_b\right\}T_c}
\end{align}
is the completely symmetric tensor of ${su(N)}$.

\section{Interaction term in the random vector formulation}\label{appB}

We will briefly outline the calculation of \eqref{layerCaction}. Using identities from appendix \ref{appA} we calculate
\begin{align}
    \strl{\Phi^4}=& \sum_{\mu_1,\ldots,\mu_4}c_1 c_2 c_3 c_4 \strl{T_1 T_2 T_3 T_4}=\\
    =&c_0^4 \frac{1}{4N}+c_0^2 6\frac{1}{4N} c_a c_a+4c_0 \frac{1}{4\sqrt{2N}} c_a c_b c_c d_{abc}\no
    &+\frac{1}{4N}\lr{\lr{c_a c_a}^2-\lr{c_a c_a}^2+\lr{c_a c_a}^2}+\frac{1}{8}
    c_a c_b c_c c_d \lr{d_{abe}d_{cde}-d_{ace}d_{bde}+d_{ade}d_{bce}}
\end{align}
The last two terms cancel due to the ${c\leftrightarrow d}$ symmetry. The first, second and fourth set of terms combine to $(\C\cdot\C)^2/4N$, since
\[(\C\cdot\C)^2=(c_0^2+c_a c_a)^2=c_0^4+(c_a c_a)^2+2 c_0^2 c_a c_a\ ,\]
and thus
\begin{align}\label{semifinal}
    \strl{\Phi^4}=\frac{1}{4N}(\C\cdot\C)^2+\frac{1}{N} c_0^2 c_a c_a +\frac{1}{\sqrt{2N}}c_0 c_a c_b c_c d_{abc}+\frac{1}{8}c_a c_b c_c c_d d_{abe}d_{cde}\ .
\end{align}
After some work, the last three terms can be written in terms of the ${N^2-1}$ matrices of size ${N^2\times N^2}$
\begin{align}
G^e=\lr{\begin{array}{cc}
     0 & v^e \\
     v^e & D^e
\end{array}}\ ,
\end{align}
where
\begin{align}
    v^e_a=\sqrt{\frac{2}{N}}\delta_{ae}\ ,\ \lr{D^e}_{ij}=d_{ije}\ ,
\end{align}
as follows
\begin{align}
    \strl{\Phi^4}=\frac{1}{4N}(\C\cdot\C)^2+\frac{1}{8}(\C\cdot G^a\cdot\C)^2\ .
\end{align}


\section{Radial derivative operators}\label{appD}
The explicit form of the $d\times d$ version of $\partial_r$ in \eqref{KR} can be expressed as follows
\begin{align}\label{partialmat}
   \mathbf{D}_R=\lr{
   {\tiny 
    \begin{array}{c|cccc|ccccccccc|cccccc}
          & \frac{1}{2\lambda} &  &  &  &  &  &  &  &  &  &  &  &  &  &  &  &  &  & \ldots  \\
         \hline
         -\frac{1}{2\lambda} &  &  &  &  & \frac{1}{2\lambda} &  &  &  &  &  &  &  &  &  &  &  &  &  & \ldots  \\
          &  &  &  &  &  & \frac{1}{2\lambda} &  &  &  &  &  &  &  &  &  &  &  &  & \ldots \\
          &  &  &  &  &  &  & \frac{1}{2\lambda} &  &  &  &  &  &  &  &  &  &  &  & \ldots  \\
          &  &  &  &  &  &  &  & \frac{1}{2\lambda} &  &  &  &  &  &  &  &  &  &  & \ldots \\
         \hline
          &-\frac{1}{2\lambda}  &  &  &  &  &  &  &  &  &  &  &  &  & \frac{1}{2\lambda} &  &  &  &  & \ldots \\
          &  &-\frac{1}{2\lambda}  &  &  &  &  &  &  &  &  &  &  &  &  & \frac{1}{2\lambda} &  &  &  & \ldots \\
          &  &  &-\frac{1}{2\lambda}  &  &  &  &  &  &  &  &  &  &  &  &  & \frac{1}{2\lambda} &  &  & \ldots \\
          &  &  &  &-\frac{1}{2\lambda}  &  &  &  &  &  &  &  &  &  &  &  & & \frac{1}{2\lambda} &  & \ldots  \\
          &  &  &  &  &  &  &  &  &  &  &  &  &  &  &  & &  & \frac{1}{2\lambda} & \ldots  \\
          &  &  &  &  &  &  &  &  &  &  &  &  &  &  &  & &  &  &  \ldots  \\
          &  &  &  &  &  &  &  &  &  &  &  &  &  &  &  & &  &  &  \ldots  \\
          &  &  &  &  &  &  &  &  &  &  &  &  &  &  &  & &  &  &  \ldots  \\
          &  &  &  &  &  &  &  &  &  &  &  &  &  &  &  & &  &  &  \ldots  \\
         \hline
          &  &  &  &  & -\frac{1}{2\lambda} &  &  &  &  &  &  &  &  &  &  & &  &  &  \ldots \\
          &  &  &  &  &  & -\frac{1}{2\lambda} &  &  &  &  &  &  &  &  &  &  &  &  &  \ldots \\
          &  &  &  &  &  &  & -\frac{1}{2\lambda} &  &  &  &  &  &  &  &  &  &  &  &  \ldots \\
          &  &  &  &  &  &  &  & -\frac{1}{2\lambda} &  &  &  &  &  &  &  &  &  &  &  \ldots \\
          &  &  &  &  &  &  &  &  & -\frac{1}{2\lambda} &  &  &  &  &  &  &  &  &  &  \ldots \\
         \vdots & \vdots & \vdots & \vdots & \vdots & \vdots & \vdots & \vdots & \vdots & \vdots & \vdots & \vdots & \vdots & \vdots & \vdots & \vdots & \vdots & \vdots & \vdots  & \ddots 
    \end{array}
    }
    }
\end{align}
and zeros are understood in the empty spaces. For the second derivative in radial direction \eqref{partialN2}, we have
\begin{align}\label{partialmat2}
   \mathbf{D}^2_R=\lr{\scalebox{.65}{$
    \begin{array}{c|cccc|ccccccccc|cccccc}
          -\frac{2}{\lambda^2}& \frac{1}{\lambda^2} &  &  &  &  &  &  &  &  &  &  &  &  &  &  &  &  &  & \ldots  \\
         \hline
         \frac{1}{\lambda^2} & -\frac{2}{\lambda^2} &  &  &  & \frac{1}{\lambda^2} &  &  &  &  &  &  &  &  &  &  &  &  &  & \ldots  \\
          &  &  -\frac{2}{\lambda^2} &  &  &  & \frac{1}{\lambda^2} &  &  &  &  &  &  &  &  &  &  &  &  & \ldots \\
          &  &  & -\frac{2}{\lambda^2} &  &  &  & \frac{1}{\lambda^2} &  &  &  &  &  &  &  &  &  &  &  & \ldots  \\
          &  &  &  & -\frac{2}{\lambda^2} &  &  &  & \frac{1}{\lambda^2} &  &  &  &  &  &  &  &  &  &  & \ldots \\
         \hline
          &\frac{1}{\lambda^2}  &  &  &  & -\frac{2}{\lambda^2} &  &  &  &  &  &  &  &  & \frac{1}{\lambda^2} &  &  &  &  & \ldots \\
          &  &\frac{1}{\lambda^2}  &  &  &  & -\frac{2}{\lambda^2} &  &  &  &  &  &  &  &  & \frac{1}{\lambda^2} &  &  &  & \ldots \\
          &  &  &\frac{1}{\lambda^2}  &  &  &  & -\frac{2}{\lambda^2} &  &  &  &  &  &  &  &  & \frac{1}{\lambda^2} &  &  & \ldots \\
          &  &  &  &\frac{1}{\lambda^2}  &  &  &  & -\frac{2}{\lambda^2} &  &  &  &  &  &  &  & & \frac{1}{\lambda^2} &  & \ldots  \\
          &  &  &  &  &  &  &  &  & -\frac{2}{\lambda^2} &  &  &  &  &  &  & &  & \frac{1}{\lambda^2} & \ldots  \\
          &  &  &  &  &  &  &  &  &  & -\frac{2}{\lambda^2} &  &  &  &  &  & &  &  &  \ldots  \\
          &  &  &  &  &  &  &  &  &  &  & -\frac{2}{\lambda^2} &  &  &  &  & &  &  &  \ldots  \\
          &  &  &  &  &  &  &  &  &  &  &  & -\frac{2}{\lambda^2} &  &  &  & &  &  &  \ldots  \\
          &  &  &  &  &  &  &  &  &  &  &  &  & -\frac{2}{\lambda^2} &  &  & &  &  &  \ldots  \\
         \hline
          &  &  &  &  & \frac{1}{\lambda^2} &  &  &  &  &  &  &  &  & -\frac{2}{\lambda^2} &  & &  &  &  \ldots \\
          &  &  &  &  &  & \frac{1}{\lambda^2} &  &  &  &  &  &  &  &  & -\frac{2}{\lambda^2} &  &  &  &  \ldots \\
          &  &  &  &  &  &  & \frac{1}{\lambda^2} &  &  &  &  &  &  &  &  & -\frac{2}{\lambda^2} &  &  &  \ldots \\
          &  &  &  &  &  &  &  & \frac{1}{\lambda^2} &  &  &  &  &  &  &  &  & -\frac{2}{\lambda^2} &  &  \ldots \\
          &  &  &  &  &  &  &  &  & \frac{1}{\lambda^2} &  &  &  &  &  &  &  &  & -\frac{2}{\lambda^2} &  \ldots \\
         \vdots & \vdots & \vdots & \vdots & \vdots & \vdots & \vdots & \vdots & \vdots & \vdots & \vdots & \vdots & \vdots & \vdots & \vdots & \vdots & \vdots & \vdots & \vdots  & \ddots 
    \end{array}$
    }}
\end{align}

\end{document}